# Direct observation of laser guided corona discharges

Tie-Jun Wang[1,*,§], Yingxia Wei[1,§], Yaoxiang Liu[1], Na Chen[1],

Yonghong Liu[1], Jingjing Ju[1], Haiyi Sun[1], Cheng Wang[1], Haihe Lu[1],

Jiansheng Liu[1,#], See Leang Chin[2], Ruxin Li[1,†], and Zhizhan Xu[1]

[1] *State Key Laboratory of High Field Laser Physics, Shanghai Institute of Optics and Fine Mechanics, Chinese Academy of Sciences, China*
[2] *centre d'Optique, Photonique et Laser (COPL) and Département de physique, de génie physique et d'optique, Université Laval, Québec, Québec G1V 0A6, Canada*

*tiejunwang@siom.ac.cn,
#michaeljs_liu@mail.siom.ac.cn,
†ruxinli@mail.shcnc.ac.cn

§these two contributed equally to the work

## Abstract

Laser based lightning control holds a promising way to solve the problem of the long standing disaster of lightning strikes. But it is a challenging project due to insufficient understanding of the interaction between laser plasma channel and high voltage electric filed. In this work, a direct observation of laser guided corona discharge is reported. The high voltage corona discharge can be guided along laser plasma filament, and enhanced through the interaction with laser filaments. The fluorescence lifetime of laser filament guided corona discharge was measured to be several microseconds, which is 3 orders of magnitude longer than the fluorescence lifetime of laser filaments. This could be advantageous towards laser assisted leader development in the atmosphere.

Lightning as a natural atmospheric discharge phenomenon is one of the long standing and most serious natural sources of disaster. Human activities toward the protection and control of lightning have never been stopped since Benjamin Franklin's famous kite experiments in 1752. Although lightning rod has been widely used to protect key locations and human life, it is impossible to avoid the lightning strikes because of its passively random nature. Active controls of lightning have been proposed and/or demonstrated. Rocket-triggered lightning discharges were successfully conducted many times in several countries since 1967 [1-6]. As a more promising and interesting method, laser based lightning control has attracted much attention [7-32] due to its pollution-free and good capability of high repetition rate operation and precise control of shooting direction.

The method of laser based lightning control is based on the plasma formation through the laser interaction with air molecules. High intensity laser pulses will ionize air molecules leading to the formation of a plasma channel required to guide lightning strikes. The idea was proposed in the 1970s [7] and investigated in the following two decades by using high energy nanosecond-duration lasers [8-10]. The ns-laser based lightning control was abandoned by the end of 1990s because of the discontinuous plasma channel formed through avalanche ionization process. The observation of femtosecond laser filamentation in 1995 [11] opened a new opportunity for laser based lightning control [12]. Femtosecond laser filamentation is a dynamic balance between intensity depended optical Kerr self-focusing and laser ionized plasma defocusing resulting in a long plasma channel formation, referred to as a laser filament [13-19]. This femtosecond laser filament with high intensity (~$5 \times 10^{13}$ W/cm$^2$ in air during free propagation) can be projected to a long distance in the atmosphere via the controls of initial pulse chirp, beam divergence etc.. The peak density of free electrons in these plasma filaments is on the order of $10^{16}$ cm$^{-3}$. However, the electron density decreases by more than an order of magnitude in less than 3 nanoseconds due to recombination process [20]. It is then followed by a much slower attachment process [21]. As a consequence, the lifetime of filament induced fluorescence in air is only a few nanoseconds, which limits the lifetime of the long plasma channel formation with high electron density. Many studies towards understanding the mechanism of filament guided discharge and developing techniques on the extension of plasma filament length and lifetime have been carried out [20-31]. The lifetime of high electron density could be extended from a few nanoseconds to several tens of nanoseconds [20-22], which is still not long enough for atmospheric lightning applications [29]. The possibility to trigger real-scale lightning in the atmosphere by laser filament has also been demonstrated although there was no direct observation of laser guided lightning strikes [29]. The observations [29, 32] suggested that corona discharges are important and may have been triggered during the interaction between laser filaments and high voltage electric field. A corona discharge is, by definition, a gas discharge where the geometry confines the gas ionizing processes to a high-field ionization region around an active electrode [33]. Corona discharge plays an important role in the leader initiation process related to lightning [34] and also find significant applications on chemical processing [35]. A direct observation of filament guided corona discharges (FGCD) and a detailed investigation on the interaction between laser filament and corona discharges are still unexplored.

In this work, we report on a direct observation of laser filament guided corona discharges. The corona discharge can be enhanced through the interaction with laser filaments. The fluorescence lifetime of laser filament guided corona discharge was measured to be several microseconds, which is 3 orders of magnitude longer than the lifetime of laser filament induced fluorescence. The results would not only benefit the understanding of the interaction between plasma filaments and corona discharges, but also stick out as the first step towards laser assisted leader development in the atmosphere.

The experiments were conducted by using a 1 kHz/25 fs Ti:sapphire chirped pulse amplification system delivering pulse energy up to 10 mJ. Laser filaments were created by a plano-convex lens with a focal length of 30 cm. The schematic of experimental setup is shown in Fig. 1. All the high voltage electric field discharge experiments were performed in a home-made Faraday cage. High voltage corona discharges were generated by using a copper electrode with a 1 mm diameter tip situated at the right hand side of Fig. 1. A DC high voltage power supply with output up to 100 kV/1000 W was connected to the electrode. (The other floated electrode at the left hand side of the figure was removed during this first set of experiments.) Laser filaments were created just next to the tip of the electrode at a distance of ~1 mm. Real color images were taken from the top at ~45 deg. to the vertical plane by a digital camera (Nikon D7000). Spectroscopic measurements of ionization-induced fluorescence were done by imaging the filament guided corona discharge channel into a CCD coupled spectrometer (Andor Shamrock SR-303i) from the side in the horizontal plane. An ICCD camera (Andor iStar 334T) and a high speed camera (PCO Dimax HD) were also used to capture the weak streamer structures of corona discharges and the temporal evolution of filament guided corona discharges from the side in the horizontal plane.

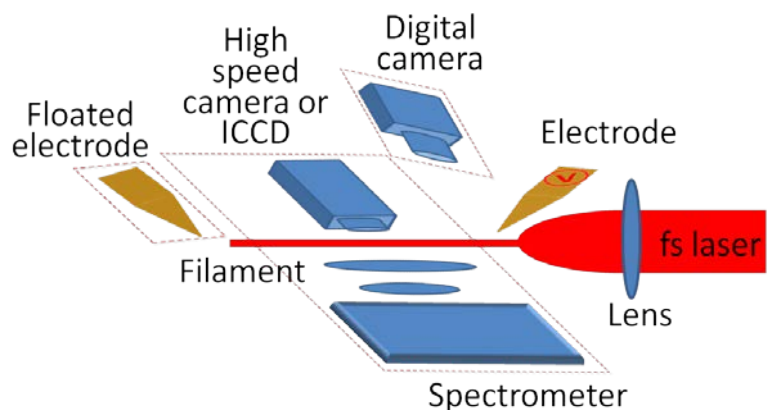

Fig. 1 the schematic of experimental setup

Real-color image of typical filament guided corona discharge is shown in Fig. 2(a), which was taken by the digital camera. The corona discharging voltage and filamenting pulse energy were set as 50 kV and 7.5 mJ, respectively. Blue color corona discharges from both ends of the filament were observed. This indicated that the corona discharges could be guided along the laser filament in or against the propagation direction. A corona discharge radiating from the tip of the electrode can also be seen. As a comparison, the real-color image of a corona discharge without laser filament is shown in Fig. 2(b) at the same voltage of 50 kV. This blue emission comes from the ionization induced UV fluorescence of air molecules (mainly from nitrogen), which will be shown in the spectral measurements later (in Fig. 6(a)). Using the ICCD camera, we observed in Fig. 2(c) the fine tree structures of the streamers corresponding to the real color discharge in the forward direction of laser propagation shown in Fig. 2(a). As a comparison, the tree structure of the streamers shown in Fig. 2(b) without the laser filament is shown in Fig. 2(d).

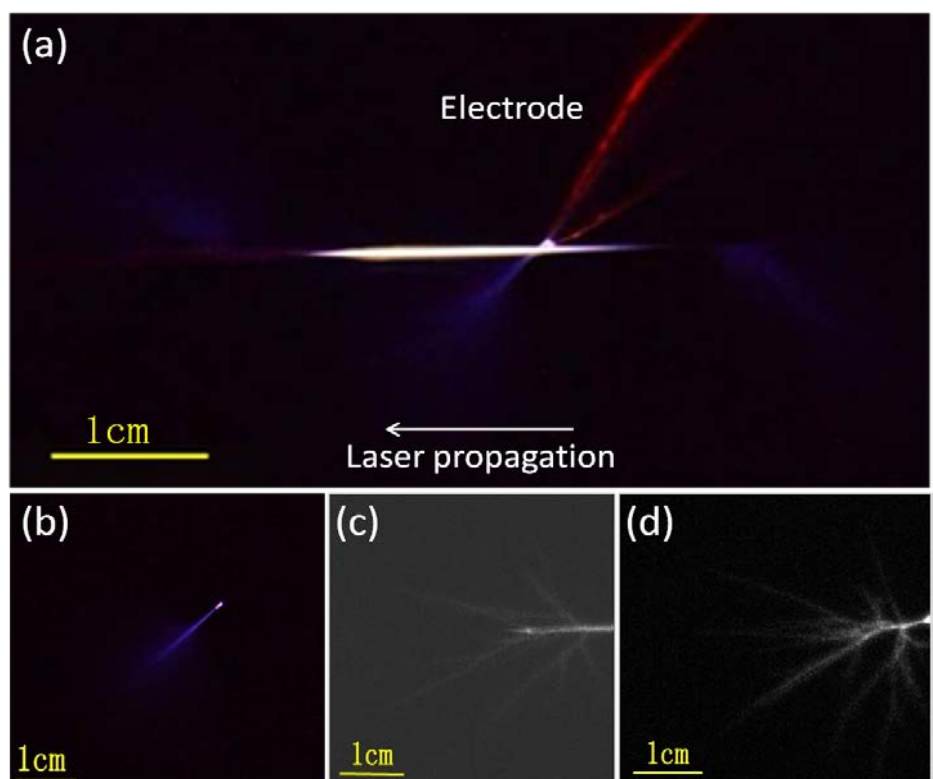

Fig. 2 (a) real-color images of typical filament guided corona discharge, (b) corona discharge without laser filament, (c) and (d) for the fine structures for those streamers in the forward direction of laser propagation from (a) and (b), respectively. The corona discharging voltage and filamenting pulse energy were 50 kV and 7.5 mJ, respectively.

To look into the corona discharge propagation along plasma filaments with different lengths, the corona discharging voltage was fixed at 50 kV. Laser filaments were created just next to the tip of the electrode at a distance of ~1 mm. Filamenting pulse energy was tuned from 0.62 mJ up to 7.83 mJ so as to generate different lengths of plasma filaments using a lens of 30 cm focal length. The guiding effect as a function of filamenting pulse energy is shown in Fig. 3. This guiding effect is sensitive to the position of initial corona discharges. When a filament is short as in the cases of 0.62 mJ and 1.31 mJ, this corona discharges can be observed at both ends of the laser filaments since the tip of the electrode is around the center of the laser filament. As the filamenting pulse energy increases, laser filaments will extend their lengths towards the focusing lens (right hand side in the figure). As a consequence, more corona discharges are generated at the leading end (right hand side) of the filament as compared to the discharges at the tailing end. The total corona discharge power was monitored by the power supply to be ~3.5 W (in Fig. 4(a)) and the fluorescence intensity of FGCD on the right hand side in Fig. 3 was obtained by integrating the pixel intensity of those images. The integrated area is indicated as a rectangle shown in the inset of Fig. 4(b). The influence of laser filament (scattering) was removed by subtracting the pixel intensity along the filament propagation direction in the area. The width of the subtracted area was 0.8 mm. The FGCD induced fluorescence signal in Fig. 4(b) increases slightly as the laser pulse energy increases. This indicates more corona discharges were guided by the laser filament, although the total corona discharge power is practically constant (Fig. 4(a)). (The jump of the last data point at the highest laser pulse energy might be due to fluctuation since no such change was observed in Fig. 4(b).)

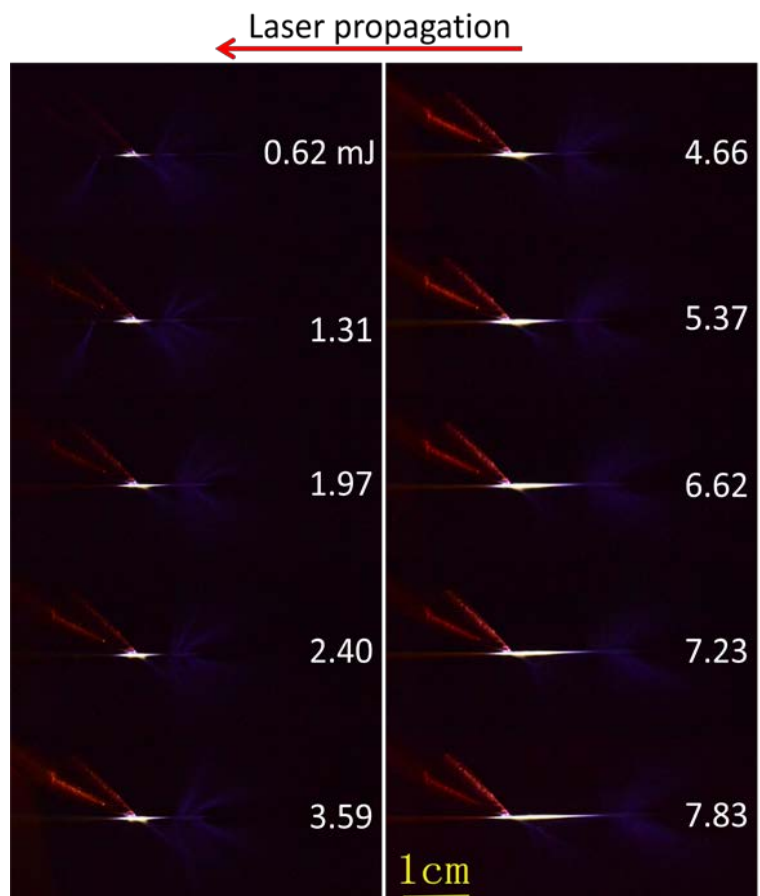

Fig. 3 Corona discharges propagation along laser filaments. Corona discharges were generated by applying a 50 kV high voltage on the electrode. Filament length was controlled by femtosecond laser pulse energy ranging from 0.62 mJ to 7.83 mJ.

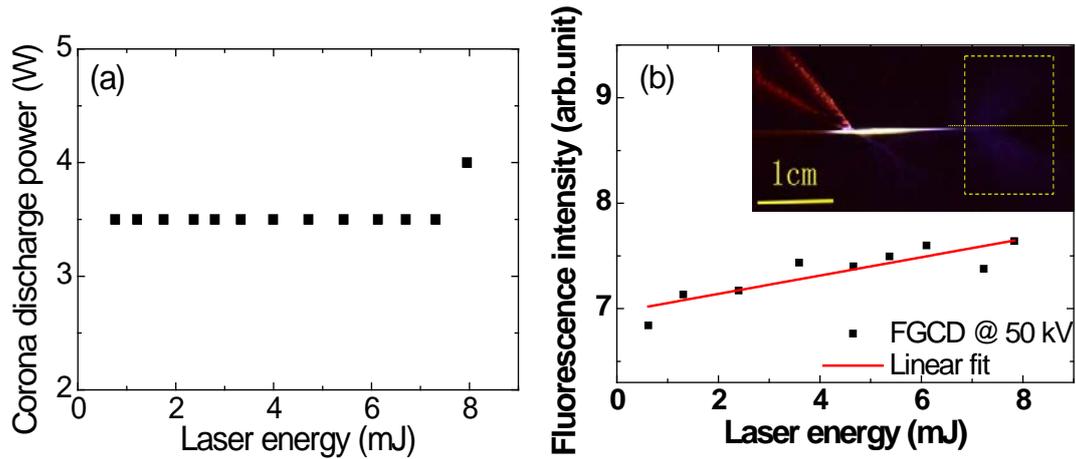

Fig. 4 Laser pulse energy dependence of corona discharge power (a) and of FGCD induced fluorescence intensity (b). The fluorescence intensity in (b) was calculated by integrating the pixel intensity in the rectangular area as shown in the inset figure from the FGCD images in Fig. 3. The intensity of laser filament in the rectangular area was removed

Fig. 5 shows a comparison of the consumed power during corona discharges with and without laser filaments. In these measurements, laser filaments were generated by focusing a 7.95 mJ laser pulse by the lens of 30 cm focal length. As the corona discharge voltage was increased, the consumed power for corona discharges exponentially increased. The corona discharges can be further increased by 10-20% through the interaction with the laser filaments, in particular, when high voltage (in our case it was >35 kV) was applied.

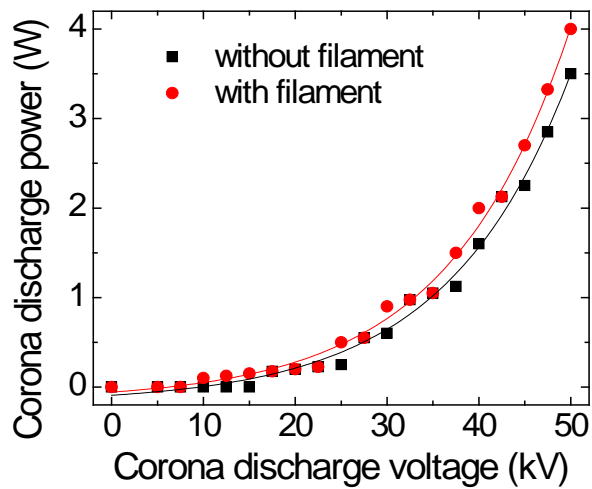

Fig. 5 Corona discharge power as a function of supplying voltage with (red round dots) and without (black square dots) laser filaments. Red and black solid lines are exponential fittings to experimental data. Filamenting pulse energy was 7.95 mJ.

This observation is confirmed by spectral measurements of the fluorescence induced by ionization processes. Fig. 6(a) depicts three typical fluorescence spectra in the UV, namely from corona discharge (CD), pure filamentation (FIL) and the plasma channel along the filament when filament guided corona discharge (FGCD) occurred. In these measurements, an identical floated electrode (left hand side in Fig. 1) was set at around 8 mm from the discharge electrode along filament propagation direction so as to easily image the discharge along the filament to the slit of the spectrometer. Adding the floated electrode would also provide one with a consistent measurement since the filament will move as the laser pulse energy is changed. A downsize imaging telescope was used to collect the fluorescence emission from the filament zone and/or the tips of the two electrodes. The voltage for the corona discharge and the laser pulse energy for filamentation in the three cases were fixed at 50 kV and 7.0 mJ, respectively. The focal length of the lens used to form a filament was 30 cm. Note that there was corona discharge spreading out from the floated electrode even without laser filament at the voltage. That emission was not collected in this measurement since attention here was paid on the filament zone under the high voltage, which would provide key information on the ionization process, hence, conducting property.

The UV spectra cover the signals from the first negative band system of $N_2^+(B^2\Sigma_u^+ - X^2\Sigma_g^+$ transition) and the second positive band system of $N_2(C^3\Pi_u - B^3\Pi_g$ transition) [36]. It is clear that the structures of the spectrum from molecular $N_2$ are similar in the three cases, but different at 391 nm and 428 nm which are ionic lines from $N_2^+$ through the transitions of $B^2\Sigma_u^+(0) - X^2\Sigma_g^+(0)$ and $B^2\Sigma_u^+(0) - X^2\Sigma_g^+(1)$. The measurements and analysis were performed in the following procedure: at a fixed voltage (50 kV) of corona discharge, the fluorescence signal as a function of filamenting pulse energy tuning range from 0.7 mJ to 7 mJ was measured; then to look into the interaction effect, the CD and FIL fluorescence signals were subtracted out from the FGCD signal. Fig. 6(b) depicts the pseudocolor plot of the resultant signal intensity as a function of the pulse energy. A clear positive resultant signal was obtained, which indicates that the stronger fluorescence came from the interaction. In particular, this fluorescence becomes much stronger when using high filamenting pulse energy (in our case when the energy was more than ~4 mJ). The enhancement of ionization induced fluorescence indicates a stronger ionization from the interaction between the laser filament and the high electric field resulting in an efficient guiding of high voltage and the enhancement of corona discharge as observed in Fig. 2-5.

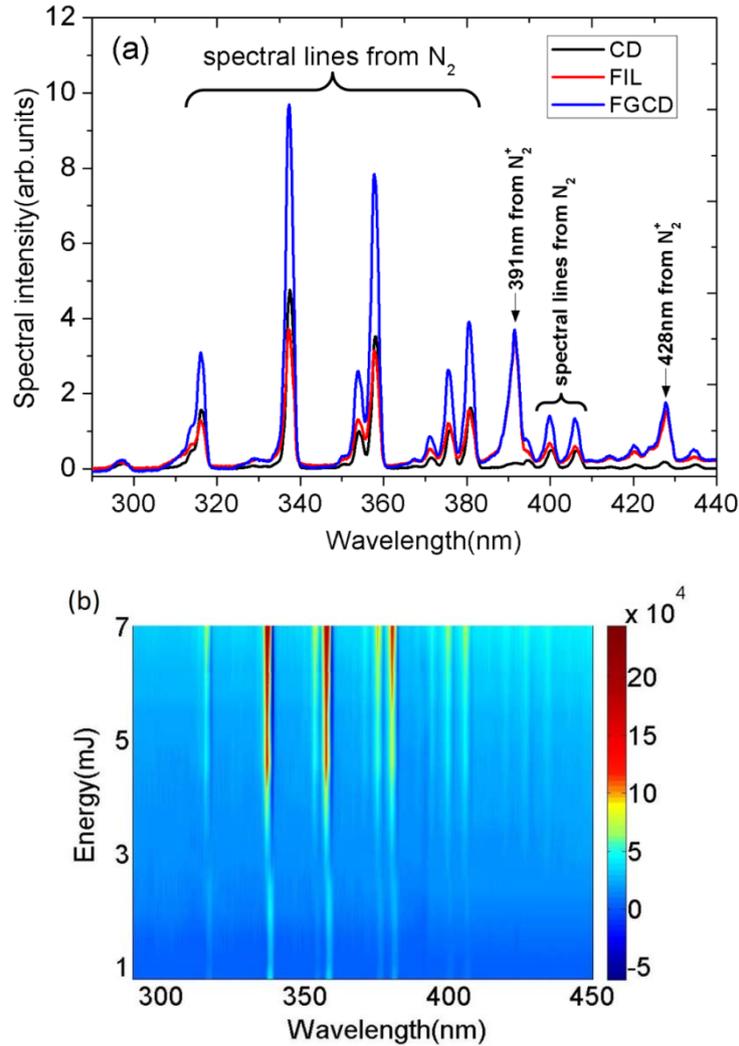

Fig. 6 (a) typical fluorescence spectrum in UV (290-440 nm) emitted by corona discharge (CD), pure filamentation (FIL) and plasma channel of filament-guided corona discharge (FGCD), respectively. The voltage for corona discharge and the laser pulse energy for filamentation in the three cases were fixed at 50 kV and 7.0 mJ, respectively. (b) pseudocolor plot of fluorescence spectral intensity of FGCD with CD and FIL fluorescence intensity subtracted as a function of the laser pulse energy tuning range from 0.7 mJ to 7 mJ. The CD voltage was fixed at 50 kV.

    The fluorescence lifetime of filament guided corona discharges were measured by using a high speed camera. High voltage and filamenting pulse energy were fixed at 50 kV and 4.8 mJ, respectively. The length of the filament focused by a lens of 30 cm focal length was ~10 mm. In order to have a clear guiding by laser filament, the tip distance of two electrodes was set at 15 mm, a little longer than the filament length. Under this condition, filament induced corona discharge could easily bridge the gap of 15 mm between the two electrodes. Note that one electrode was floated in order to maintain the corona discharge. The exposure time of the high speed camera was 1 μs, which was triggered by the 1 kHz laser pulses. The triggering time was delayed with respect to the laser arrival time. At each delay time, hundreds of

shots were recorded. When filament guided corona discharge bridges the two electrodes, it was counted as a successful one. The probability of successful filament guided corona discharges is shown in Fig. 7. It indicates that this guided discharge can last up to ~4 μs, which is 3 orders of magnitude longer than the fluorescence lifetime of the plasma filament. This observation may solve the long standing problem of short lifetime of plasma filament for atmospheric lightning.

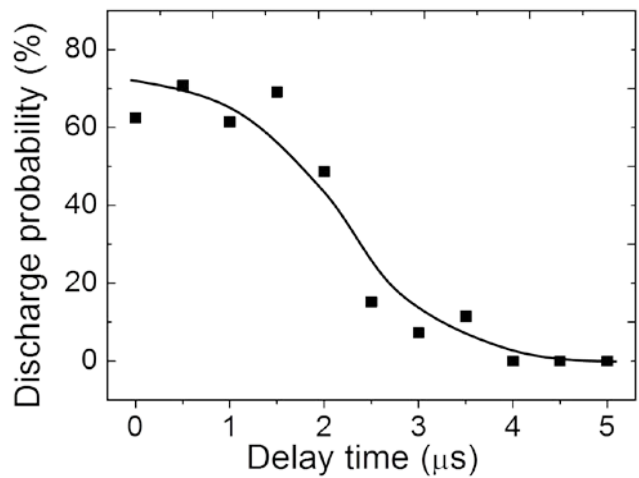

Fig. 7 Lifetime of filament guided corona discharges. Filamenting pulse energy was 4.8 mJ. Corona discharging voltage was 50 kV.

The main mechanisms involved in the laser filament guided corona discharges include the photo-ionization, impact ionization, electron attachment on oxygen molecules and the detachment of these electron by various processes such as electron collision, ion collision, etc., charged particle recombination, electron diffusion etc. [24,27]. The high intensity inside the laser filament not only ionizes air molecules (mostly $N_2$ and $O_2$), but also excites molecular ions and neutrals into high lying states resulting in fluorescence emission of neutrals and ions [13-15,37]. The long low density plasma channel is the key to guide high voltage and to form the corona discharge at the two tips of the filament. The filament is firstly heated by the Joule effect after electron-ion recombination and then hydrodynamically expands outward resulting in a low pressure in the filament relaxation zone [38,39]. The lower pressure enhances the electrical conductivity because of less collision of electrons. The laser initiated electrons together with corona discharged electrons undergo impact ionization resulting in a higher plasma density. As a consequence, corona discharges are enhanced (as seen by corona discharge power measurement in Fig. 5) , which is also confirmed by the spectral measurement of filament guided corona discharge (FGCD) induced fluorescence in Fig. 6.

The longer decay time of the FGCD could be understood as follows. During FGCD, photo-ionization is an ultrafast process, which occurs within the duration of femtosecond laser pulse. The attachment of electrons onto oxygen molecules is detrimental to the lifetime of the plasma channel. When the external electric field is added along the plasma channel, the detachments of electrons from oxygen molecules will be increased, leading to a much longer decay time in the FGCD channel as was observed in Fig. 7 [21].

In summary, laser filament guided corona discharges were directly observed. The corona discharge can be enhanced under the condition of single or multiple filamentation. The filament guided corona discharges possess the good properties of long lifetime and the possibility for long distance propagation, which are critical and important for the applications of atmospheric lightning control.

**Acknowledgments**
This work was supported in part by National Natural Science Foundation of China (Grant Nos. 61221064, 11127901), National 973 Project (Grant No. 2011CB808103), Chinese Academy of Sciences and the State Key Laboratory of High Field Laser Physics, 100 Talents Program of Chinese Academy of Sciences, Shanghai Pujiang Program. The authors acknowledge LUSTER LightTech (Beijing) Co. Ltd. for providing the high speed camera.

**Additional information**

**Competing Financial Interests statement:** The authors declare no competing financial interests.